\documentclass[%
superscriptaddress,
 amsmath,amssymb,
 aps,
 prl,
floatfix,
notitlepage,
11pt
]{revtex4-2}

\usepackage{graphicx}
\usepackage{dcolumn}
\usepackage{bm}
\usepackage{xcolor}

\begin{document}


\title{Coupling and Decoupling of Polaritonic States in Multimode Cavities}

\author{M Balasubrahmaniyam}
\affiliation{
 School of Chemistry, Raymond and Beverly Sackler Faculty of Exact Sciences and Tel Aviv University Center for Light-Matter Interaction, Tel Aviv University, Tel Aviv 6997801, Israel.} 

\author{Cyriaque Genet}
\affiliation{
 Universit\'e de Strasbourg, CNRS, Institut de Science et d\textsc{\char13}Ing\'enierie Supramol\'eculaires, UMR 7006, 67000 Strasbourg, France
}
 
\author{Tal Schwartz}
 \email{talschwartz@tau.ac.il}
\affiliation{
 School of Chemistry, Raymond and Beverly Sackler Faculty of Exact Sciences and Tel Aviv University Center for Light-Matter Interaction, Tel Aviv University, Tel Aviv 6997801, Israel.}



\begin{abstract}
\vspace{20mm}
We demonstrate a new type of transition within the strong coupling regime, which alters the coupling mechanism in multimode cavities. We show that this transition drastically modifies the Hamiltonian describing the polaritons, such that different cavity modes are either entangled via the material or completely decoupled. This decoupling transition occurs due to the competition between the dissipation in the material and the finite group velocity, which governs the propagation of information across the cavity and among the molecules. The results indicate that the velocity of light, which is often not taken into account in cavity quantum electrodynamics, plays a crucial role in the formation of cavity polaritons and their dynamics.
\end{abstract}

\maketitle

\newpage
In cavity quantum electrodynamics, the balance between the light-matter coupling strength, which is quantified by the Rabi frequency $\Omega$, and the other time-scales in the system defines several different regimes~\cite{Haroche2010}.
In weak coupling, irreversible processes dominate the dynamics and the cavity acts as a perturbation, which may modify the spontaneous emission rate, among other effects.
When the coupling strength overcomes the incoherent processes, the system crosses into the strong coupling regime, in which the coherent mixing between the material excitations and photonic wavefunctions confined by the cavity gives rise to hybrid light-matter excitations known as cavity polaritons.
This regime is marked by an energetic splitting~$\hbar\Omega$ of the resonantly-coupled states of the material and the optical cavity.
Strong coupling had been studied in many different systems, such as atomic cavities~\cite{Raimond2001}, inorganic semiconductors~\cite{Skolnick1998} and molecules~\cite{Ebbesen2016a}, and it offer numerous opportunities for optoelectronic and quantum technologies~\cite{Sanvitto2016,Raimond2001}.
When the coupling is further increased and becomes comparable to the transition energy in the material, a different regime, known as ultrastrong coupling, is reached~\cite{Ciuti2005,Schwartz2011,Scalari2012,Kena-Cohen2013} and new phenomena emerge, such as squeezed vacuum and emission of entangled photon pairs~\cite{FriskKockum2019}.

Recently, the possibility of hybridizing multiple transitions to a cavity mode has been explored, primarily in organic systems~\cite{Lidzey2000,Holmes2004,Hakala2009,Muallem2016,Menghrajani2019,Hernandez2019,Schutz2020}. In such systems, a middle polariton branch may emerges as a coherent tripartite superposition of the photonic wavefunction and two material excitons. As such, these hybrid polaritonic excitations give rise to indirect energy transfer and other effects~\cite{Zhong2016,Balasubrahmaniyam2017,Reitz2018,Hertzog2019,Lather2019} which are particularly relevant in molecular chemistry and material science~\cite{Ebbesen2016a}.

In this work, we consider the scenario in which a single excitonic transition is coupled to a multimode Fabry-P\'erot cavity~\cite{Coles2014,George2016a,Li2020}. We show that in such multimode cavities, where several optical resonances can interact with the material, a new kind of transition occurs within the strong coupling regime, with the system crossing between two fundamentally different coupling mechanisms. As we show, this transition results from the competition between the lifetime of the collectively coupled emitters and the finite propagation velocity of light circulating inside the cavity. 

The cavity is composed of two parallel mirrors, with the gap between them uniformly filled with an optically-active material (e.g. quantum wells or molecules). If the cavity length $L$ is large enough, the cavity free spectral range, given by $\Delta\nu=c/{2Ln_0}$ (with $c$ being the speed of light, and $n_0$ the background refractive index inside the cavity), can be small enough such that several longitudinal modes reside in the vicinity of the transition energy (denoted as $E_x$) and multiple cavity modes may therefore effectively couple with the material excitation. Similar conditions can  also be obtained in other geometries, such as ring resonators or dielectric microspheres supporting multiple whispering gallery modes~\cite{Vasista2020}.

\begin{figure}
    \includegraphics[width=120mm]{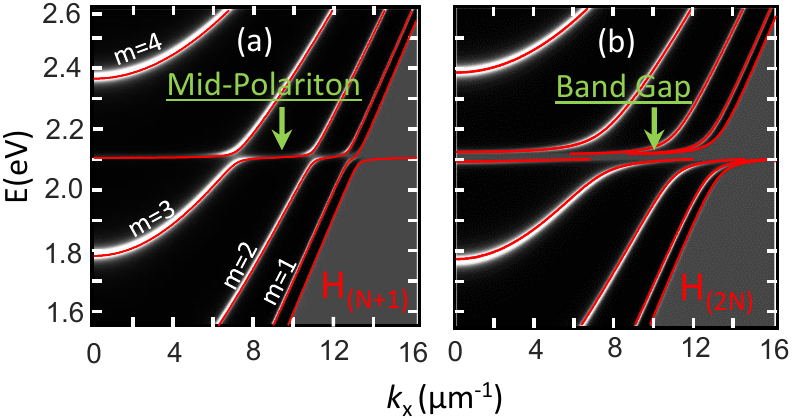}
    \caption{Reflectively (gray-scale image, calculated for transverse electric polarization) of a hybrid cavity under strong coupling with $f=0.01 eV^2$ (a) and $f=0.1 eV^2$ (b). The low reflectively (bright regions) corresponds to the polariton branches. The red lines show a fit to the polariton energies calculated using the Hamiltonian $H_{N+1}$ (a) and $H_{2N}$ (b) with $g=20$ and $70$ meV, respectively. In constructing both Hamiltonians, the energies of the uncoupled cavity modes were obtained from the TMM simulations using $f=0$).}
    \label{fig:disp}
\end{figure}

To simulate the polariton dispersion of the system, we used the transfer matrix method (TMM)~\cite{George2016a,Saleh1991} to calculate its reflection spectrum as a function of the in-plane momentum $k_x$, as shown in Fig.~\ref{fig:disp}. In these simulations, the mirrors are defined as silver layers of 45 nm thickness and the material is modeled as a Lorentz-type dispersive medium, with a refractive index given by $n(E)=\left[n_0^2+ {f}/{(E_x^2-E^2-i\gamma E)}\right]^{1/2}$.
Here $\gamma$ is the FWHM of the transition (or the decay rate) and $f$ is the oscillator strength (per unit volume) of the transition. The oscillator strength is related to the transition dipole element $d$ by $f=2N E_x d^2/n_0^2$~\cite{Torma2015}, where $N$ is the density of emitters, and it determines the collective coupling strength in the system, which is given by $\Omega=\sqrt{f}/n_0$ (see Ref.~\cite{Agranovich2003} and Supplemental Material). We choose the parameters $E_x=2.11~\text{eV}$,~$\gamma=10~\text{meV}$ and ~$n_0=1.5$, matching typical parameters for TDBC J-aggregate molecules, which are commonly used in strong coupling experiments~\cite{Bellessa2004,Zengin2013,Rozenman2018}. Fig.~\ref{fig:disp}(a) and~\ref{fig:disp}(b) present the simulated dispersion with oscillator strengths of $0.01~eV^2$ and $0.1~eV^2$, respectively. The cavity length was set to $L=700$ nm, such that the $1^{st}$, $2^{nd}$ and $3^{rd}$ order modes are resonant with the exciton at $k_x =$ 12.5, 10.5 and 6.5  $\mu m ^{-1}$. Fig.~\ref{fig:disp}(a) clearly shows the avoided crossings behavior typical of strong coupling at these resonant points, along with a slight blue shift of the fourth-order mode, which is not resonant with the exciton for any value of $k_x$. In addition, the dispersion also exhibits a "mid-polariton" that adiabatically transforms from one cavity mode to the next, passing continuously through the exciton energy. By analogy with a two exciton, single cavity-mode system~\cite{Lidzey2000}, this system can be modeled using an extended coupled oscillator model, which is given by an $N+1$ dimensional Hamiltonian in the form
\begin{equation}
H_{N+1}=
\begin{pmatrix}
\tilde{E_x} & g & g & \dots &g\\
g & {E_c^{(1)}} & 0 \\
g & 0 & {E_c^{(2)}} \\
\vdots& & &   \ddots& 0\\
g & & & 0 &  {E_c^{(N)}}
\end{pmatrix}_{(N+1)\times(N+1)}
\end{equation}
where $\tilde{E_x}=E_x+i\gamma/2$, ${E_c^{(n)}}(k_x)$ are the resonant energies of the $N$ dispersive photonic modes of the cavity and the $2N$ off-diagonal elements $g=\hbar\Omega/2$  couple the exciton to each one of the cavity modes~\cite{Coles2014}. The red lines in Fig.~\ref{fig:disp}(a) show the fit of the diagonalized Hamiltonian $H_{N+1}$ (with four cavity modes) to the simulation results, using $g=20$ meV as a single fitting parameter.

In sharp contrast, at a higher value of the oscillator strength, [Fig.~\ref{fig:disp}(b), $f=0.1~eV^2$], along with the expected increase in the Rabi-splitting energy (which scales as $\sqrt{f}$), the dispersion clearly shows a gap spanning the limited k-space from $k_x = 0$ and up to the light line, and separating $N$ upper and $N$ lower polaritonic branches. This energy gap is reminiscent of the polaritonic bandgap discussed in the context of ultrastrong coupling~\cite{Askenazi2014,George2016a}, even though our system operates within the usual strong coupling regime. Since there are $2N$ polaritonic modes, the system clearly cannot be described by the $H_{N+1}$. Instead, we find that the dispersion can be fitted by considering the independent coupling of each one of the cavity modes to the excitonic transition, which is described by the $2N$-dimensional block-diagonal Hamiltonian
\begin{equation}
    H_{2N}=
    \begin{pmatrix}
    {E_c^{(1)}} & g & 0& 0 & \dots &0 &0  \\
    g & \tilde{E_x} & 0 & 0\\
    0 & 0 & {E_c^{(2)}} & g \\
    0 & 0 & g & \tilde{E_x} \\
    \vdots & & & & \ddots\\
    0 & & & & & {E_c^{(N)}} & g \\
    0 & & & & &  g & \tilde{E_x} \\
    \end{pmatrix}_{2N\times2N}
\end{equation}
such that each $2\times 2$ block represents the interaction between the excitonic mode and a specific cavity mode. The solutions of this Hamiltonian (calculated using $g=70$ meV), are plotted in Fig.~\ref{fig:disp}(b) (red lines), perfectly matching the results of the TMM simulations. It is clear that in this regime the exciton-mediated coupling between the cavity modes, which occurs at the lower coupling strength [Fig.~\ref{fig:disp}(a)], completely disappears, and each one of the $N$ pairs of polaritonic branches is associated with a distinct cavity mode.

Since two different Hamiltonians with different dimensionality, $H_{N+1}$ and $H_{2N}$, are required for modeling the system at two different coupling strengths, our results clearly indicate that there is a transition within the strong coupling regime, for which the nature of the coupling changes fundamentally. This complex behavior is different from all previously known transitions for light-matter interaction in a cavity~\footnote{For example, in the case of the transition from weak coupling to strong coupling, the transition can be described by a non-hermitian Hamiltonian which accounts for the loss rates, and these completely determine the interaction strength for the transition happens (i.e. when the energy degeneracy is removed).
The transition we observe here is also clearly different from the crossover to the ultra-strong coupling regime, as the strong-coupling Hamiltonian is simply an approximate representation of the ultrastrong coupling Hamiltonian under the rotating-wave approximation. In contrast, since the structures of $H_{N+1}$ and $H_{2N}$ are completely different, neither one of them can be obtained from the other in any limit.
Moreover, it can be shown that the behavior seen in Fig.~\ref{fig:disp} is reproduced even when a full Hamiltonian, which includes the counter-rotating terms (which is typically used under ultrastrong coupling conditions), is used (see Supplemental Material).}.

\begin{figure}[b]
    \includegraphics[width=120mm]{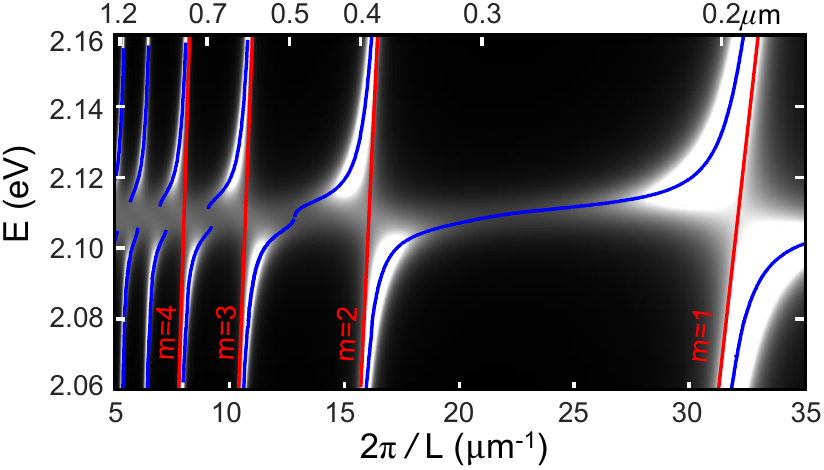}
    \caption{Reflection spectrum as a function of inverse cavity length (a measure of longitudinal momentum), calculated at normal incidence and with~$f=0.015~eV^2$. Low reflectance (bright regions) corresponds to the polariton branches. The blue lines show the reflection minima, depicting the evolution of the polaritons with the thickness of the cavity.}
    \label{fig:withL}
\end{figure}

To further investigate the observed transition, we use the TMM simulations to obtain the dispersion at an intermediate coupling strength, setting $f=0.015~eV^2$ and scanning over the cavity length from $0.180~\mu m $ to $ 1.25~ \mu m $ (while maintaining the same conditions of a uniformly-filled cavity with a fixed material concentration).
Fig.~\ref{fig:withL} presents the simulated normal-incidence reflection spectrum as a function of inverse cavity length.
As the cavity length is varied, different cavity modes become resonant with the material, causing a normal mode splitting of each one of the cavity mode branches, as seen earlier in Fig.~\ref{fig:disp}.
Between these anti-crossing points, the results clearly show a transition from a regime in which a continuous mid-polaritons branch exists (e.g. around $25~\mu m^{-1}$ or $L\simeq0.25~\mu m$) to a regime in which the mid-polaritons disappear and a gap is formed between the polaritonic branches (below $10~\mu m^{-1}$ or $L>0.630$~$\mu$m). 
It should be noted that the Rabi splitting is identical for all the cavity modes, as observed at the avoided crossing points, irrespective of cavity length (or the mode volume).
This is expected since, for a homogeneously filled cavity, the coupling strength depends only on the oscillator strength per unit volume, which scales linearly with the density of emitters.
On the other hand, it is clear from Fig.~\ref{fig:withL} that the behavior of the strongly coupled system is governed by the dimensions of the cavity, and not only by the coupling strength and loss rates.
Once again, the observed dependence on the cavity length also indicates that the physics of the coupled system cannot be fully captured by either one of the Hamiltonians above.

\begin{figure}[t]
\includegraphics[width=120mm]{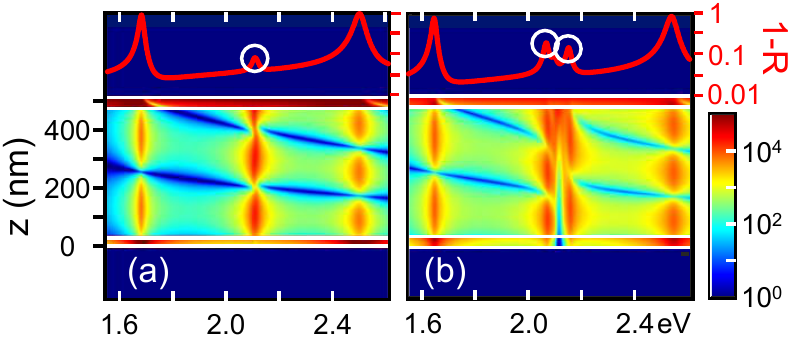}
\caption{Spatial distribution of the energy dissipation within the cavity as a function of the photon energy, calculated for normal incidence (from the top side) with a cavity length $L\simeq 0.490~\mu\text{m}$ and $f=0.01~\text{eV}^2$ (a) and $f=0.1~\text{eV}^2$ (b). The white lines mark the edges of the mirror. Also shown is the spectral response ($1-R$, red curves) with the circles marking the position of the mid-polariton mode in (a) and the decoupled polaritonic modes in (b).}
\label{fig:Iofx}
\end{figure}

To understand the nature of the mid-polaritons and their decoupling across the transition, we examine their spatial-spectral distribution before and after the transition.
Fig.~\ref{fig:Iofx} shows the calculated energy loss distribution within the cavity~\cite{Schwartz2011}, which is defined as $\frac{1}{2}Im\{n^2\}\omega\left|\mathcal{E}(z)/\mathcal{E}_0\right|^2$, where $\omega$ is the frequency, $\mathcal{E}(z)$ describes the field distribution inside the cavity (extracted from the TMM simulations) and $\mathcal{E}_0$ is the incoming field amplitude.
The data were calculated for normal incidence and under similar conditions to Fig.~\ref{fig:disp}, with the cavity length ($L\simeq 0.490\mu m$)  chosen such that the second- and third-order modes are equally detuned (in opposite directions) from the excitonic transition.
As seen, below the transition [Fig.~\ref{fig:Iofx}(a), $f=0.01~\text{eV}^2$], the energy distribution at 1.68 and 2.5~eV exhibits standing-wave patterns with two and three antinodes, respectively, corresponding to photon-like polaritonic states associated with the second- and third-order modes.
At the mid-polariton energy ($2.11$~eV) the data also displays a standing-wave pattern. However, its distribution does not correspond to an integer number of cycles within the cavity, as one may expect for any mode of the system.
Interestingly, under such conditions the accumulated round-trip phase at that energy is $5\pi$, which usually (i.e. in the absence of the active medium) gives rise to destructive interference between the mirrors and full reflection from the cavity.
However, when the excitonic material is placed inside the cavity, this destructive interference is diminished, leading to the creation of the mid-polariton state.
With a higher coupling strength [Fig.~\ref{fig:Iofx}(b), $f=0.1~\text{eV}^2$] and above the transition, we again observe the splitting of the mid-polariton peak in two, while the peaks at 1.68~eV and 2.5~eV remain unchanged (apart for a small shift further away from $E_x$).
Moreover, the energy loss distribution clearly shows that the splitting of the mid-polariton results in two separate polaritonic modes having different structures.
The left one (2.07~eV) has three antinodes, and therefore it is associated with the third-order cavity mode, while the right one (at 2.15~eV) has two antinodes, linking it to the second-order mode.
Therefore, these results demonstrate how the increase in the coupling strength transforms the system from a regime in which the two cavity modes are mixed via their mutual coupling with the material (as described by $H_{N+1}$), to a regime in which the transition in the material couples separately to each one of the the two cavity modes as described by $H_{2N}$, giving rise to two decoupled pairs of polaritonic states.

Next, in order to elucidate the nature of the observed transition, we monitor the emergence of the mid-polariton by looking at the intersection between the accumulated polariton round-trip phase for the multiple branches using $\phi_p=2 E_p(L) n(E_p)L/{\hbar c}$, calculated from the polariton energies $E_p(L)$~\footnote{Since close to the transition neither one of the hamiltonians can be used, the length-dependent polariton energies used for calculating the polaritonic phase were obtained from the TMM simulations, in a similar manner to Fig.~\ref{fig:withL}}, and the photonic round trip phase $\phi_{RT}(E)=2k(E)L$, calculated with a fixed cavity length and with the energy taken as a free parameter.
Such an approach has been used previously as a semiclassical model for predicting the emergence of strong coupling and the polariton energies (see Supplemental Material and Refs.~\cite{Zhu1990,George2016a}).
Interestingly, as we show in the Supplemental Material, at the transition to strong coupling, the group velocity for photons circulating within the cavity (which is directly related to the slope of $\phi_{RT}$) crosses from a positive to a negative value.
\begin{figure}[t]
    \includegraphics[width=120mm]{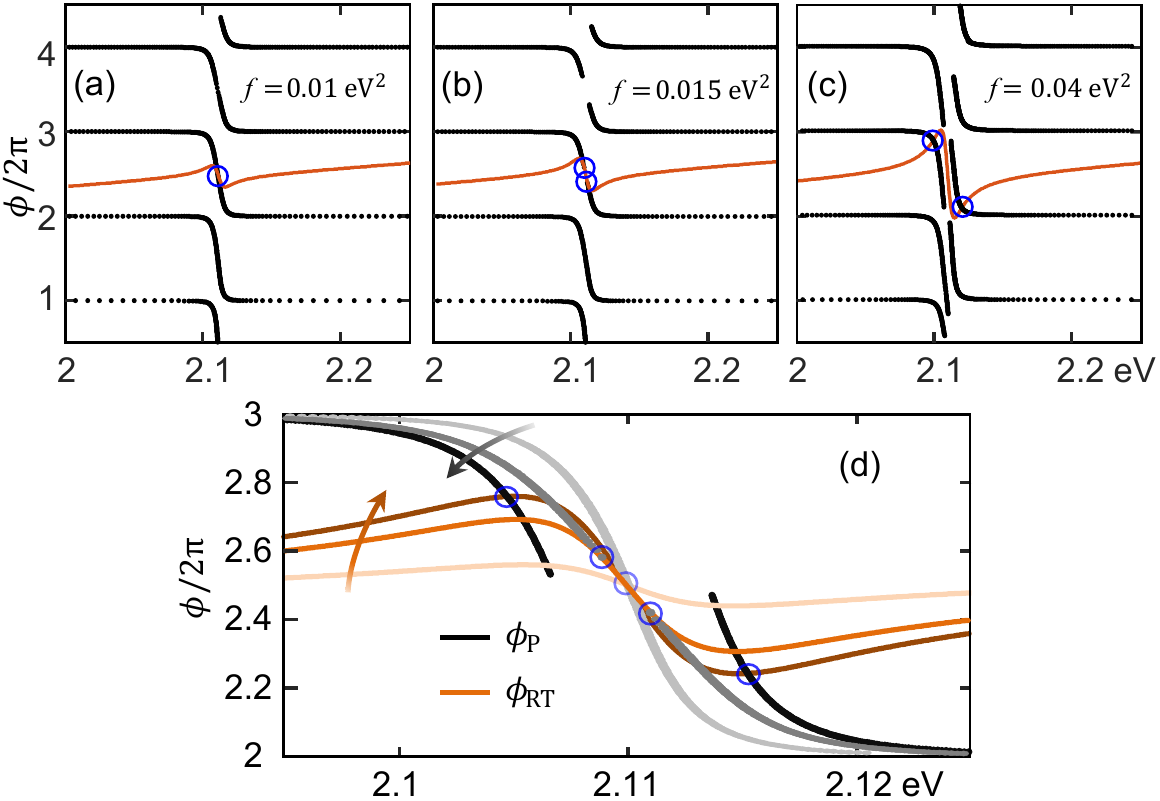}
    \caption{(a)-(c) Accumulated polaritonic phase $\phi_p$ with a varying cavity length (black dotted lines) and photonic round-trip phase $\phi_{RT}$ calculated for $L=$0.490~$\mu$m and with the energy as a free parameter (red lines). The blue circles mark the hybrid modes at the intersection points for $L=$0.490~$\mu$m. (d) Phase curves for increasing values of $f$ (indicated by the arrows) in the vicinity of the transition, plotted near the mid-polariton residing between the $2^{nd}$ and $3^{rd}$ modes.}
    \label{fig:wide}
\end{figure}
In Fig.~\ref{fig:wide}(a)-(c) we plot the polaritonic phase $\phi_p$ as a function of $E_p$ (black curves) for cavity lengths between $0.1$ and $0.9~\mu\text{m}$ and with $f=0.01$, $ 0.015 $ and $0.04~eV^2$.
For a cavity length of $L=490$ nm (for which $E_x$ is located exactly between the $2^{nd}$ and $3^{rd}$-order cavity modes) these $f$ values correspond to the system being below, near, or above the coupling-decoupling transition.
As can be seen, far from the absorption line (where $n(E)$ is almost purely real), $\phi_p$ is a multiple integer of $2\pi$, as expected.
However, close to $E_x$, the polaritonic phase crosses continuously from $3 \times 2\pi$ to $2 \times 2\pi$ as the cavity thickness decreases and the polariton energy increases, passing through a value of $5\pi$ at $L\sim 490$ nm as also seen in Fig.~\ref{fig:Iofx}(b).
In addition, we overlay the photonic phase $\phi_{RT}(E)=2k(E)L$ for this specific choice of $L$ [red curve in Figs.~\ref{fig:wide}(a)-(c)].
For each value of $f$, the intersection points between these curves and $\phi_p$, marked by blue circles, give the polariton energy for this particular cavity length.
When the value of $f$ increases [Figs.~\ref{fig:wide}(b),(c)], we observe that the polariton phase curve acquires a discontinuity as the mid-polariton branch splits and the two decoupled polaritons are formed.
At the same time, since the dispersion of the excitonic medium depends linearly on $f$, as $f$ increases, the slope of $\phi_{RT}$ (red curves) at $E=E_x$ becomes more negative.
In examining the region close to the transition point [see Fig.~\ref{fig:wide}(d)], we find that the decoupling between the cavity modes and the appearance of a gap occur exactly at the point where the two slopes become equal, which is a necessary requirement for having two intersection points between the curves.
This observation indicates that the coupling-decoupling transition occurs when the (negative) group velocity crosses a critical value, which, as shown in the Supplemental Material, is given by $\left|v_g\right|=\frac{\pi L \gamma}{h}(1-2\beta)$, where $\beta=e^{-\alpha L}/(1-e^{-\alpha L})$ is a small parameter expressing the absorbance within the cavity at $E_x$ and $\alpha \simeq 2/L+{2\pi n_0 \gamma}/{hc}$ is the absorption coefficient of the medium. Furthermore, for a long enough cavity, $\beta$ can be neglected such that the critical group velocity is simply given by $\left|v_g\right|=\pi L\gamma/h$.

\begin{figure}[t]
    \includegraphics[width=120mm]{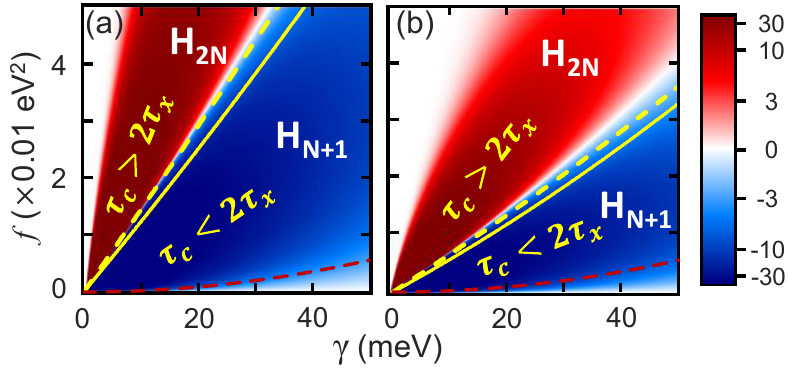}
    \caption{Coupling-decoupling transition in the $(\gamma,f)$ parameter space. The false-color image shows the exact numerical solution for $\frac{\partial^2 R}{\partial E^2}\big|_{E_x}$ calculated for (a) $m=2$ and (b) $m=5$,  and the yellow curves correspond to the approximated solution for the critical group velocity with $\beta=0$ (solid) or evaluated for $\beta=e^{-\alpha L}/(1-e^{-\alpha L})$. The dashed red line marks the transition to the strong coupling regime.}
    \label{fig:PT}
\end{figure}

This result can be interpreted in the following manner: the time required for a photon to travel across the cavity is given by $\tau_c=L/|v_g|$. On the other hand, the exciton lifetime is given by $\tau_x=\frac{h}{2\pi\gamma}$.
With these, we identify that the decoupling transition and the splitting of the mid-polariton branch occurs when $\tau_c > 2\tau_x$, that is, when the dissipation within the material occurs before the photon that couples the emitters with the cavity can traverse the entire volume of the cavity.
In order to examine the validity of this result, we may numerically evaluate the second derivative of the reflection spectrum at $E=E_x$, where a positive value corresponds to a single reflection dip of a mid-polariton mode while a negative value indicates the splitting of the mid-polariton into two decoupled states.
This is shown in Fig.~\ref{fig:PT}, where the cavity length was chosen such that a mid-polariton branch crosses between the second and third cavity modes [Fig.~\ref{fig:PT}(a), $L=490$~nm] or between the fifth and sixth mode [Fig.~\ref{fig:PT}(b), $L=1077$~nm].
Superimposed on this exact calculation, we plot the analytical result based on the expression for $\left|v_g\right|$ with $\beta=0$ (solid line) and with $\beta\neq 0$ approximated as above (dashed lines).
As a reference, we also plot the curve corresponding to the transition from weak to strong coupling, given by $f=n_0^2 \gamma^2$.
As can be seen, the analytical solution agrees well with the exact calculation, with only a slight influence upon the inclusion of $\beta$.   

To conclude, we presented the study of strong light-matter coupling in multimode cavities and showed that such systems exhibit a transition between two different types of strong coupling behaviors.
Below the transition, different cavity modes interact via the material while above the transition the cavity modes decouple and the material interacts individually with each cavity mode.
Our analysis shows that the observed transition between two different regimes of strong coupling is related to the competition between the finite group velocity of photons in the cavity and the decay rate of the emitters, indicating that spatiotemporal effects play an important role in the dynamics of strong coupling.
Consequently, it should be expected that systems with similar coupling strength but different dimensions can exhibit different coherent dynamics.
Moreover, the group velocity may be further tailored, by including either passive or nonlinear dispersive media within a cavity in addition to the excitonic material.
In this way, the exciton-mediated coupling between the cavity modes can be controlled, which may be used as a tool for various quantum information processing schemes~\cite{Dong2009,Prado2011,Wickenbrock2013}.
Finally, since neither one of the Hamiltonians, $H_{N+1}$ nor $H_{2N}$, can be used to describe the cavity in both coupling regimes, it is evident from our findings that a new type of Hamiltonian, which takes into account the finite rate of information spreading among the emitters, is required for capturing the complete physics of strongly coupled systems.

\begin{acknowledgments}
    This work was supported in part by the Israel Science Foundation, grant no. 1435/19, Agence Nationale de la Recherche (ANR), France, ANR Equipex Union (Grant No. ANR-10-EQPX-52-01), the Labex NIE projects (Grant No. ANR-11-LABX-0058-NIE), and USIAS within the Investissements d'Avenir program (Grant No. ANR-10-IDEX-0002-02). M.B. is grateful for the financial support from Tel Aviv University under the Raymond and Beverly Sackler Post-Doctoral Scholarship and from the Council for higher Education, Israel, under the PBC Postdoctoral Fellowship.
\end{acknowledgments}

\bibliography{CouplingDecoupling}

\begin{thebibliography}{39}%
\makeatletter
\providecommand \@ifxundefined [1]{%
 \@ifx{#1\undefined}
}%
\providecommand \@ifnum [1]{%
 \ifnum #1\expandafter \@firstoftwo
 \else \expandafter \@secondoftwo
 \fi
}%
\providecommand \@ifx [1]{%
 \ifx #1\expandafter \@firstoftwo
 \else \expandafter \@secondoftwo
 \fi
}%
\providecommand \natexlab [1]{#1}%
\providecommand \enquote  [1]{``#1''}%
\providecommand \bibnamefont  [1]{#1}%
\providecommand \bibfnamefont [1]{#1}%
\providecommand \citenamefont [1]{#1}%
\providecommand \href@noop [0]{\@secondoftwo}%
\providecommand \href [0]{\begingroup \@sanitize@url \@href}%
\providecommand \@href[1]{\@@startlink{#1}\@@href}%
\providecommand \@@href[1]{\endgroup#1\@@endlink}%
\providecommand \@sanitize@url [0]{\catcode `\\12\catcode `\$12\catcode
  `\&12\catcode `\#12\catcode `\^12\catcode `\_12\catcode `\%12\relax}%
\providecommand \@@startlink[1]{}%
\providecommand \@@endlink[0]{}%
\providecommand \url  [0]{\begingroup\@sanitize@url \@url }%
\providecommand \@url [1]{\endgroup\@href {#1}{\urlprefix }}%
\providecommand \urlprefix  [0]{URL }%
\providecommand \Eprint [0]{\href }%
\providecommand \doibase [0]{https://doi.org/}%
\providecommand \selectlanguage [0]{\@gobble}%
\providecommand \bibinfo  [0]{\@secondoftwo}%
\providecommand \bibfield  [0]{\@secondoftwo}%
\providecommand \translation [1]{[#1]}%
\providecommand \BibitemOpen [0]{}%
\providecommand \bibitemStop [0]{}%
\providecommand \bibitemNoStop [0]{.\EOS\space}%
\providecommand \EOS [0]{\spacefactor3000\relax}%
\providecommand \BibitemShut  [1]{\csname bibitem#1\endcsname}%
\let\auto@bib@innerbib\@empty
\bibitem [{\citenamefont {Haroche}\ and\ \citenamefont
  {Raimond}(2010)}]{Haroche2010}%
  \BibitemOpen
  \bibfield  {author} {\bibinfo {author} {\bibfnamefont {S.}~\bibnamefont
  {Haroche}}\ and\ \bibinfo {author} {\bibfnamefont {J.~M.}\ \bibnamefont
  {Raimond}},\ }\href
  {https://doi.org/10.1093/acprof:oso/9780198509141.001.0001} {\emph {\bibinfo
  {title} {Exploring the Quantum: Atoms, Cavities, and Photons}}}\ (\bibinfo
  {publisher} {Oxford University Press},\ \bibinfo {year} {2010})\ pp.\
  \bibinfo {pages} {1--616}\BibitemShut {NoStop}%
\bibitem [{\citenamefont {Raimond}\ \emph {et~al.}(2001)\citenamefont
  {Raimond}, \citenamefont {Brune},\ and\ \citenamefont
  {Haroche}}]{Raimond2001}%
  \BibitemOpen
  \bibfield  {author} {\bibinfo {author} {\bibfnamefont {J.~M.}\ \bibnamefont
  {Raimond}}, \bibinfo {author} {\bibfnamefont {M.}~\bibnamefont {Brune}},\
  and\ \bibinfo {author} {\bibfnamefont {S.}~\bibnamefont {Haroche}},\
  }\bibfield  {title} {\bibinfo {title} {{Colloquium: Manipulating quantum
  entanglement with atoms and photons in a cavity}},\ }\href
  {https://doi.org/10.1103/RevModPhys.73.565} {\bibfield  {journal} {\bibinfo
  {journal} {Reviews of Modern Physics}\ }\textbf {\bibinfo {volume} {73}},\
  \bibinfo {pages} {565} (\bibinfo {year} {2001})}\BibitemShut {NoStop}%
\bibitem [{\citenamefont {Skolnick}\ \emph {et~al.}(1998)\citenamefont
  {Skolnick}, \citenamefont {Fisher},\ and\ \citenamefont
  {Whittaker}}]{Skolnick1998}%
  \BibitemOpen
  \bibfield  {author} {\bibinfo {author} {\bibfnamefont {M.~S.}\ \bibnamefont
  {Skolnick}}, \bibinfo {author} {\bibfnamefont {T.~A.}\ \bibnamefont
  {Fisher}},\ and\ \bibinfo {author} {\bibfnamefont {D.~M.}\ \bibnamefont
  {Whittaker}},\ }\bibfield  {title} {\bibinfo {title} {{Strong coupling
  phenomena in quantum microcavity structures}},\ }\href
  {https://doi.org/10.1088/0268-1242/13/7/003} {\bibfield  {journal} {\bibinfo
  {journal} {Semiconductor Science and Technology}\ }\textbf {\bibinfo {volume}
  {13}},\ \bibinfo {pages} {645} (\bibinfo {year} {1998})}\BibitemShut
  {NoStop}%
\bibitem [{\citenamefont {Ebbesen}(2016)}]{Ebbesen2016a}%
  \BibitemOpen
  \bibfield  {author} {\bibinfo {author} {\bibfnamefont {T.~W.}\ \bibnamefont
  {Ebbesen}},\ }\bibfield  {title} {\bibinfo {title} {{Hybrid Light–Matter
  States in a Molecular and Material Science Perspective}},\ }\href
  {https://doi.org/10.1021/acs.accounts.6b00295} {\bibfield  {journal}
  {\bibinfo  {journal} {Accounts of Chemical Research}\ }\textbf {\bibinfo
  {volume} {49}},\ \bibinfo {pages} {2403} (\bibinfo {year}
  {2016})}\BibitemShut {NoStop}%
\bibitem [{\citenamefont {Sanvitto}\ and\ \citenamefont
  {K{\'{e}}na-Cohen}(2016)}]{Sanvitto2016}%
  \BibitemOpen
  \bibfield  {author} {\bibinfo {author} {\bibfnamefont {D.}~\bibnamefont
  {Sanvitto}}\ and\ \bibinfo {author} {\bibfnamefont {S.}~\bibnamefont
  {K{\'{e}}na-Cohen}},\ }\bibfield  {title} {\bibinfo {title} {{The road
  towards polaritonic devices}},\ }\href {https://doi.org/10.1038/nmat4668}
  {\bibfield  {journal} {\bibinfo  {journal} {Nature Materials}\ }\textbf
  {\bibinfo {volume} {15}},\ \bibinfo {pages} {1061} (\bibinfo {year}
  {2016})}\BibitemShut {NoStop}%
\bibitem [{\citenamefont {Ciuti}\ \emph {et~al.}(2005)\citenamefont {Ciuti},
  \citenamefont {Bastard},\ and\ \citenamefont {Carusotto}}]{Ciuti2005}%
  \BibitemOpen
  \bibfield  {author} {\bibinfo {author} {\bibfnamefont {C.}~\bibnamefont
  {Ciuti}}, \bibinfo {author} {\bibfnamefont {G.}~\bibnamefont {Bastard}},\
  and\ \bibinfo {author} {\bibfnamefont {I.}~\bibnamefont {Carusotto}},\
  }\bibfield  {title} {\bibinfo {title} {{Quantum vacuum properties of the
  intersubband cavity polariton field}},\ }\href
  {https://doi.org/10.1103/PhysRevB.72.115303} {\bibfield  {journal} {\bibinfo
  {journal} {Physical Review B}\ }\textbf {\bibinfo {volume} {72}},\ \bibinfo
  {pages} {115303} (\bibinfo {year} {2005})}\BibitemShut {NoStop}%
\bibitem [{\citenamefont {Schwartz}\ \emph {et~al.}(2011)\citenamefont
  {Schwartz}, \citenamefont {Hutchison}, \citenamefont {Genet},\ and\
  \citenamefont {Ebbesen}}]{Schwartz2011}%
  \BibitemOpen
  \bibfield  {author} {\bibinfo {author} {\bibfnamefont {T.}~\bibnamefont
  {Schwartz}}, \bibinfo {author} {\bibfnamefont {J.~A.}\ \bibnamefont
  {Hutchison}}, \bibinfo {author} {\bibfnamefont {C.}~\bibnamefont {Genet}},\
  and\ \bibinfo {author} {\bibfnamefont {T.~W.}\ \bibnamefont {Ebbesen}},\
  }\bibfield  {title} {\bibinfo {title} {{Reversible Switching of Ultrastrong
  Light-Molecule Coupling}},\ }\href
  {https://doi.org/10.1103/PhysRevLett.106.196405} {\bibfield  {journal}
  {\bibinfo  {journal} {Physical Review Letters}\ }\textbf {\bibinfo {volume}
  {106}},\ \bibinfo {pages} {196405} (\bibinfo {year} {2011})}\BibitemShut
  {NoStop}%
\bibitem [{\citenamefont {Scalari}\ \emph {et~al.}(2012)\citenamefont
  {Scalari}, \citenamefont {Maissen}, \citenamefont {Turcinkova}, \citenamefont
  {Hagenmuller}, \citenamefont {{De Liberato}}, \citenamefont {Ciuti},
  \citenamefont {Reichl}, \citenamefont {Schuh}, \citenamefont {Wegscheider},
  \citenamefont {Beck},\ and\ \citenamefont {Faist}}]{Scalari2012}%
  \BibitemOpen
  \bibfield  {author} {\bibinfo {author} {\bibfnamefont {G.}~\bibnamefont
  {Scalari}}, \bibinfo {author} {\bibfnamefont {C.}~\bibnamefont {Maissen}},
  \bibinfo {author} {\bibfnamefont {D.}~\bibnamefont {Turcinkova}}, \bibinfo
  {author} {\bibfnamefont {D.}~\bibnamefont {Hagenmuller}}, \bibinfo {author}
  {\bibfnamefont {S.}~\bibnamefont {{De Liberato}}}, \bibinfo {author}
  {\bibfnamefont {C.}~\bibnamefont {Ciuti}}, \bibinfo {author} {\bibfnamefont
  {C.}~\bibnamefont {Reichl}}, \bibinfo {author} {\bibfnamefont
  {D.}~\bibnamefont {Schuh}}, \bibinfo {author} {\bibfnamefont
  {W.}~\bibnamefont {Wegscheider}}, \bibinfo {author} {\bibfnamefont
  {M.}~\bibnamefont {Beck}},\ and\ \bibinfo {author} {\bibfnamefont
  {J.}~\bibnamefont {Faist}},\ }\bibfield  {title} {\bibinfo {title}
  {{Ultrastrong Coupling of the Cyclotron Transition of a 2D Electron Gas to a
  THz Metamaterial}},\ }\href {https://doi.org/10.1126/science.1216022}
  {\bibfield  {journal} {\bibinfo  {journal} {Science}\ }\textbf {\bibinfo
  {volume} {335}},\ \bibinfo {pages} {1323} (\bibinfo {year}
  {2012})}\BibitemShut {NoStop}%
\bibitem [{\citenamefont {K{\'{e}}na-Cohen}\ \emph {et~al.}(2013)\citenamefont
  {K{\'{e}}na-Cohen}, \citenamefont {Maier},\ and\ \citenamefont
  {Bradley}}]{Kena-Cohen2013}%
  \BibitemOpen
  \bibfield  {author} {\bibinfo {author} {\bibfnamefont {S.}~\bibnamefont
  {K{\'{e}}na-Cohen}}, \bibinfo {author} {\bibfnamefont {S.~a.}\ \bibnamefont
  {Maier}},\ and\ \bibinfo {author} {\bibfnamefont {D.~D.~C.}\ \bibnamefont
  {Bradley}},\ }\bibfield  {title} {\bibinfo {title} {{Ultrastrongly Coupled
  Exciton-Polaritons in Metal-Clad Organic Semiconductor Microcavities}},\
  }\href {https://doi.org/10.1002/adom.201300256} {\bibfield  {journal}
  {\bibinfo  {journal} {Advanced Optical Materials}\ }\textbf {\bibinfo
  {volume} {1}},\ \bibinfo {pages} {827} (\bibinfo {year} {2013})}\BibitemShut
  {NoStop}%
\bibitem [{\citenamefont {{Frisk Kockum}}\ \emph {et~al.}(2019)\citenamefont
  {{Frisk Kockum}}, \citenamefont {Miranowicz}, \citenamefont {{De Liberato}},
  \citenamefont {Savasta},\ and\ \citenamefont {Nori}}]{FriskKockum2019}%
  \BibitemOpen
  \bibfield  {author} {\bibinfo {author} {\bibfnamefont {A.}~\bibnamefont
  {{Frisk Kockum}}}, \bibinfo {author} {\bibfnamefont {A.}~\bibnamefont
  {Miranowicz}}, \bibinfo {author} {\bibfnamefont {S.}~\bibnamefont {{De
  Liberato}}}, \bibinfo {author} {\bibfnamefont {S.}~\bibnamefont {Savasta}},\
  and\ \bibinfo {author} {\bibfnamefont {F.}~\bibnamefont {Nori}},\ }\bibfield
  {title} {\bibinfo {title} {{Ultrastrong coupling between light and matter}},\
  }\href {https://doi.org/10.1038/s42254-018-0006-2} {\bibfield  {journal}
  {\bibinfo  {journal} {Nature Reviews Physics}\ }\textbf {\bibinfo {volume}
  {1}},\ \bibinfo {pages} {19} (\bibinfo {year} {2019})}\BibitemShut {NoStop}%
\bibitem [{\citenamefont {Lidzey}\ \emph {et~al.}(2000)\citenamefont {Lidzey},
  \citenamefont {Bradley}, \citenamefont {Armitage}, \citenamefont {Walker},\
  and\ \citenamefont {Skolnick}}]{Lidzey2000}%
  \BibitemOpen
  \bibfield  {author} {\bibinfo {author} {\bibfnamefont {D.~G.}\ \bibnamefont
  {Lidzey}}, \bibinfo {author} {\bibfnamefont {D.~D.~C.}\ \bibnamefont
  {Bradley}}, \bibinfo {author} {\bibfnamefont {A.}~\bibnamefont {Armitage}},
  \bibinfo {author} {\bibfnamefont {S.}~\bibnamefont {Walker}},\ and\ \bibinfo
  {author} {\bibfnamefont {M.~S.}\ \bibnamefont {Skolnick}},\ }\bibfield
  {title} {\bibinfo {title} {{Photon-Mediated Hybridization of Frenkel Excitons
  in Organic Semiconductor Microcavities}},\ }\href
  {https://doi.org/10.1126/science.288.5471.1620} {\bibfield  {journal}
  {\bibinfo  {journal} {Science}\ }\textbf {\bibinfo {volume} {288}},\ \bibinfo
  {pages} {1620} (\bibinfo {year} {2000})}\BibitemShut {NoStop}%
\bibitem [{\citenamefont {Holmes}\ and\ \citenamefont
  {Forrest}(2004)}]{Holmes2004}%
  \BibitemOpen
  \bibfield  {author} {\bibinfo {author} {\bibfnamefont {R.~J.}\ \bibnamefont
  {Holmes}}\ and\ \bibinfo {author} {\bibfnamefont {S.~R.}\ \bibnamefont
  {Forrest}},\ }\bibfield  {title} {\bibinfo {title} {{Strong Exciton-Photon
  Coupling and Exciton Hybridization in a Thermally Evaporated Polycrystalline
  Film of an Organic Small Molecule}},\ }\href
  {https://doi.org/10.1103/PhysRevLett.93.186404} {\bibfield  {journal}
  {\bibinfo  {journal} {Physical Review Letters}\ }\textbf {\bibinfo {volume}
  {93}},\ \bibinfo {pages} {186404} (\bibinfo {year} {2004})}\BibitemShut
  {NoStop}%
\bibitem [{\citenamefont {Hakala}\ \emph {et~al.}(2009)\citenamefont {Hakala},
  \citenamefont {Toppari}, \citenamefont {Kuzyk}, \citenamefont {Pettersson},
  \citenamefont {Tikkanen}, \citenamefont {Kunttu},\ and\ \citenamefont
  {T{\"{o}}rm{\"{a}}}}]{Hakala2009}%
  \BibitemOpen
  \bibfield  {author} {\bibinfo {author} {\bibfnamefont {T.~K.}\ \bibnamefont
  {Hakala}}, \bibinfo {author} {\bibfnamefont {J.~J.}\ \bibnamefont {Toppari}},
  \bibinfo {author} {\bibfnamefont {A.}~\bibnamefont {Kuzyk}}, \bibinfo
  {author} {\bibfnamefont {M.}~\bibnamefont {Pettersson}}, \bibinfo {author}
  {\bibfnamefont {H.}~\bibnamefont {Tikkanen}}, \bibinfo {author}
  {\bibfnamefont {H.}~\bibnamefont {Kunttu}},\ and\ \bibinfo {author}
  {\bibfnamefont {P.}~\bibnamefont {T{\"{o}}rm{\"{a}}}},\ }\bibfield  {title}
  {\bibinfo {title} {{Vacuum rabi splitting and strong-coupling dynamics for
  surface-plasmon polaritons and rhodamine 6G molecules}},\ }\href
  {https://doi.org/10.1103/PhysRevLett.103.053602} {\bibfield  {journal}
  {\bibinfo  {journal} {Physical Review Letters}\ }\textbf {\bibinfo {volume}
  {103}},\ \bibinfo {pages} {053602} (\bibinfo {year} {2009})}\BibitemShut
  {NoStop}%
\bibitem [{\citenamefont {Muallem}\ \emph {et~al.}(2016)\citenamefont
  {Muallem}, \citenamefont {Palatnik}, \citenamefont {Nessim},\ and\
  \citenamefont {Tischler}}]{Muallem2016}%
  \BibitemOpen
  \bibfield  {author} {\bibinfo {author} {\bibfnamefont {M.}~\bibnamefont
  {Muallem}}, \bibinfo {author} {\bibfnamefont {A.}~\bibnamefont {Palatnik}},
  \bibinfo {author} {\bibfnamefont {G.~D.}\ \bibnamefont {Nessim}},\ and\
  \bibinfo {author} {\bibfnamefont {Y.~R.}\ \bibnamefont {Tischler}},\
  }\bibfield  {title} {\bibinfo {title} {{Strong Light-Matter Coupling and
  Hybridization of Molecular Vibrations in a Low-Loss Infrared Microcavity}},\
  }\href {https://doi.org/10.1021/acs.jpclett.6b00617} {\bibfield  {journal}
  {\bibinfo  {journal} {The Journal of Physical Chemistry Letters}\ }\textbf
  {\bibinfo {volume} {7}},\ \bibinfo {pages} {2002} (\bibinfo {year}
  {2016})}\BibitemShut {NoStop}%
\bibitem [{\citenamefont {Menghrajani}\ \emph {et~al.}(2019)\citenamefont
  {Menghrajani}, \citenamefont {Fernandez}, \citenamefont {Nash},\ and\
  \citenamefont {Barnes}}]{Menghrajani2019}%
  \BibitemOpen
  \bibfield  {author} {\bibinfo {author} {\bibfnamefont {K.~S.}\ \bibnamefont
  {Menghrajani}}, \bibinfo {author} {\bibfnamefont {H.~A.}\ \bibnamefont
  {Fernandez}}, \bibinfo {author} {\bibfnamefont {G.~R.}\ \bibnamefont
  {Nash}},\ and\ \bibinfo {author} {\bibfnamefont {W.~L.}\ \bibnamefont
  {Barnes}},\ }\bibfield  {title} {\bibinfo {title} {{Hybridization of Multiple
  Vibrational Modes via Strong Coupling Using Confined Light Fields}},\ }\href
  {https://doi.org/10.1002/adom.201900403} {\bibfield  {journal} {\bibinfo
  {journal} {Advanced Optical Materials}\ }\textbf {\bibinfo {volume} {7}},\
  \bibinfo {pages} {1900403} (\bibinfo {year} {2019})}\BibitemShut {NoStop}%
\bibitem [{\citenamefont {Hern{\'{a}}ndez}\ and\ \citenamefont
  {Herrera}(2019)}]{Hernandez2019}%
  \BibitemOpen
  \bibfield  {author} {\bibinfo {author} {\bibfnamefont {F.~J.}\ \bibnamefont
  {Hern{\'{a}}ndez}}\ and\ \bibinfo {author} {\bibfnamefont {F.}~\bibnamefont
  {Herrera}},\ }\bibfield  {title} {\bibinfo {title} {{Multi-level quantum Rabi
  model for anharmonic vibrational polaritons}},\ }\href
  {https://doi.org/10.1063/1.5121426} {\bibfield  {journal} {\bibinfo
  {journal} {The Journal of Chemical Physics}\ }\textbf {\bibinfo {volume}
  {151}},\ \bibinfo {pages} {144116} (\bibinfo {year} {2019})}\BibitemShut
  {NoStop}%
\bibitem [{\citenamefont {Sch{\"{u}}tz}\ \emph {et~al.}(2020)\citenamefont
  {Sch{\"{u}}tz}, \citenamefont {Schachenmayer}, \citenamefont
  {Hagenm{\"{u}}ller}, \citenamefont {Brennen}, \citenamefont {Volz},
  \citenamefont {Sandoghdar}, \citenamefont {Ebbesen}, \citenamefont {Genes},\
  and\ \citenamefont {Pupillo}}]{Schutz2020}%
  \BibitemOpen
  \bibfield  {author} {\bibinfo {author} {\bibfnamefont {S.}~\bibnamefont
  {Sch{\"{u}}tz}}, \bibinfo {author} {\bibfnamefont {J.}~\bibnamefont
  {Schachenmayer}}, \bibinfo {author} {\bibfnamefont {D.}~\bibnamefont
  {Hagenm{\"{u}}ller}}, \bibinfo {author} {\bibfnamefont {G.~K.}\ \bibnamefont
  {Brennen}}, \bibinfo {author} {\bibfnamefont {T.}~\bibnamefont {Volz}},
  \bibinfo {author} {\bibfnamefont {V.}~\bibnamefont {Sandoghdar}}, \bibinfo
  {author} {\bibfnamefont {T.~W.}\ \bibnamefont {Ebbesen}}, \bibinfo {author}
  {\bibfnamefont {C.}~\bibnamefont {Genes}},\ and\ \bibinfo {author}
  {\bibfnamefont {G.}~\bibnamefont {Pupillo}},\ }\bibfield  {title} {\bibinfo
  {title} {{Ensemble-Induced Strong Light-Matter Coupling of a Single Quantum
  Emitter}},\ }\href {https://doi.org/10.1103/physrevlett.124.113602}
  {\bibfield  {journal} {\bibinfo  {journal} {Physical Review Letters}\
  }\textbf {\bibinfo {volume} {124}},\ \bibinfo {pages} {113602} (\bibinfo
  {year} {2020})}\BibitemShut {NoStop}%
\bibitem [{\citenamefont {Zhong}\ \emph {et~al.}(2016)\citenamefont {Zhong},
  \citenamefont {Chervy}, \citenamefont {Wang}, \citenamefont {George},
  \citenamefont {Thomas}, \citenamefont {Hutchison}, \citenamefont {Devaux},
  \citenamefont {Genet},\ and\ \citenamefont {Ebbesen}}]{Zhong2016}%
  \BibitemOpen
  \bibfield  {author} {\bibinfo {author} {\bibfnamefont {X.}~\bibnamefont
  {Zhong}}, \bibinfo {author} {\bibfnamefont {T.}~\bibnamefont {Chervy}},
  \bibinfo {author} {\bibfnamefont {S.}~\bibnamefont {Wang}}, \bibinfo {author}
  {\bibfnamefont {J.}~\bibnamefont {George}}, \bibinfo {author} {\bibfnamefont
  {A.}~\bibnamefont {Thomas}}, \bibinfo {author} {\bibfnamefont {J.~A.}\
  \bibnamefont {Hutchison}}, \bibinfo {author} {\bibfnamefont {E.}~\bibnamefont
  {Devaux}}, \bibinfo {author} {\bibfnamefont {C.}~\bibnamefont {Genet}},\ and\
  \bibinfo {author} {\bibfnamefont {T.~W.}\ \bibnamefont {Ebbesen}},\
  }\bibfield  {title} {\bibinfo {title} {{Non-Radiative Energy Transfer
  Mediated by Hybrid Light-Matter States}},\ }\href
  {https://doi.org/10.1002/anie.201600428} {\bibfield  {journal} {\bibinfo
  {journal} {Angewandte Chemie International Edition}\ }\textbf {\bibinfo
  {volume} {55}},\ \bibinfo {pages} {6202} (\bibinfo {year}
  {2016})}\BibitemShut {NoStop}%
\bibitem [{\citenamefont {Balasubrahmaniyam}\ \emph {et~al.}(2017)\citenamefont
  {Balasubrahmaniyam}, \citenamefont {Kar}, \citenamefont {Sen}, \citenamefont
  {Bisht},\ and\ \citenamefont {Kasiviswanathan}}]{Balasubrahmaniyam2017}%
  \BibitemOpen
  \bibfield  {author} {\bibinfo {author} {\bibfnamefont {M.}~\bibnamefont
  {Balasubrahmaniyam}}, \bibinfo {author} {\bibfnamefont {D.}~\bibnamefont
  {Kar}}, \bibinfo {author} {\bibfnamefont {P.}~\bibnamefont {Sen}}, \bibinfo
  {author} {\bibfnamefont {P.~B.}\ \bibnamefont {Bisht}},\ and\ \bibinfo
  {author} {\bibfnamefont {S.}~\bibnamefont {Kasiviswanathan}},\ }\bibfield
  {title} {\bibinfo {title} {{Observation of subwavelength localization of
  cavity plasmons induced by ultra-strong exciton coupling}},\ }\href
  {https://doi.org/10.1063/1.4979838} {\bibfield  {journal} {\bibinfo
  {journal} {Applied Physics Letters}\ }\textbf {\bibinfo {volume} {110}},\
  \bibinfo {pages} {171101} (\bibinfo {year} {2017})}\BibitemShut {NoStop}%
\bibitem [{\citenamefont {Reitz}\ \emph {et~al.}(2018)\citenamefont {Reitz},
  \citenamefont {Mineo},\ and\ \citenamefont {Genes}}]{Reitz2018}%
  \BibitemOpen
  \bibfield  {author} {\bibinfo {author} {\bibfnamefont {M.}~\bibnamefont
  {Reitz}}, \bibinfo {author} {\bibfnamefont {F.}~\bibnamefont {Mineo}},\ and\
  \bibinfo {author} {\bibfnamefont {C.}~\bibnamefont {Genes}},\ }\bibfield
  {title} {\bibinfo {title} {{Energy transfer and correlations in
  cavity-embedded donor-acceptor configurations}},\ }\href
  {https://doi.org/10.1038/s41598-018-27396-z} {\bibfield  {journal} {\bibinfo
  {journal} {Scientific Reports}\ }\textbf {\bibinfo {volume} {8}},\ \bibinfo
  {pages} {9050} (\bibinfo {year} {2018})}\BibitemShut {NoStop}%
\bibitem [{\citenamefont {Hertzog}\ \emph {et~al.}(2019)\citenamefont
  {Hertzog}, \citenamefont {Wang}, \citenamefont {Mony},\ and\ \citenamefont
  {B{\"{o}}rjesson}}]{Hertzog2019}%
  \BibitemOpen
  \bibfield  {author} {\bibinfo {author} {\bibfnamefont {M.}~\bibnamefont
  {Hertzog}}, \bibinfo {author} {\bibfnamefont {M.}~\bibnamefont {Wang}},
  \bibinfo {author} {\bibfnamefont {J.}~\bibnamefont {Mony}},\ and\ \bibinfo
  {author} {\bibfnamefont {K.}~\bibnamefont {B{\"{o}}rjesson}},\ }\bibfield
  {title} {\bibinfo {title} {{Strong light–matter interactions: a new
  direction within chemistry}},\ }\href {https://doi.org/10.1039/C8CS00193F}
  {\bibfield  {journal} {\bibinfo  {journal} {Chemical Society Reviews}\
  }\textbf {\bibinfo {volume} {48}},\ \bibinfo {pages} {937} (\bibinfo {year}
  {2019})}\BibitemShut {NoStop}%
\bibitem [{\citenamefont {Lather}\ \emph {et~al.}(2019)\citenamefont {Lather},
  \citenamefont {Bhatt}, \citenamefont {Thomas}, \citenamefont {Ebbesen},\ and\
  \citenamefont {George}}]{Lather2019}%
  \BibitemOpen
  \bibfield  {author} {\bibinfo {author} {\bibfnamefont {J.}~\bibnamefont
  {Lather}}, \bibinfo {author} {\bibfnamefont {P.}~\bibnamefont {Bhatt}},
  \bibinfo {author} {\bibfnamefont {A.}~\bibnamefont {Thomas}}, \bibinfo
  {author} {\bibfnamefont {T.~W.}\ \bibnamefont {Ebbesen}},\ and\ \bibinfo
  {author} {\bibfnamefont {J.}~\bibnamefont {George}},\ }\bibfield  {title}
  {\bibinfo {title} {{Cavity Catalysis by Cooperative Vibrational Strong
  Coupling of Reactant and Solvent Molecules}},\ }\href
  {https://doi.org/10.1002/anie.201905407} {\bibfield  {journal} {\bibinfo
  {journal} {Angewandte Chemie International Edition}\ }\textbf {\bibinfo
  {volume} {58}},\ \bibinfo {pages} {10635} (\bibinfo {year}
  {2019})}\BibitemShut {NoStop}%
\bibitem [{\citenamefont {Coles}\ and\ \citenamefont
  {Lidzey}(2014)}]{Coles2014}%
  \BibitemOpen
  \bibfield  {author} {\bibinfo {author} {\bibfnamefont {D.~M.}\ \bibnamefont
  {Coles}}\ and\ \bibinfo {author} {\bibfnamefont {D.~G.}\ \bibnamefont
  {Lidzey}},\ }\bibfield  {title} {\bibinfo {title} {{A ladder of polariton
  branches formed by coupling an organic semiconductor exciton to a series of
  closely spaced cavity-photon modes}},\ }\href
  {https://doi.org/10.1063/1.4876604} {\bibfield  {journal} {\bibinfo
  {journal} {Applied Physics Letters}\ }\textbf {\bibinfo {volume} {104}},\
  \bibinfo {pages} {191108} (\bibinfo {year} {2014})}\BibitemShut {NoStop}%
\bibitem [{\citenamefont {George}\ \emph {et~al.}(2016)\citenamefont {George},
  \citenamefont {Chervy}, \citenamefont {Shalabney}, \citenamefont {Devaux},
  \citenamefont {Hiura}, \citenamefont {Genet},\ and\ \citenamefont
  {Ebbesen}}]{George2016a}%
  \BibitemOpen
  \bibfield  {author} {\bibinfo {author} {\bibfnamefont {J.}~\bibnamefont
  {George}}, \bibinfo {author} {\bibfnamefont {T.}~\bibnamefont {Chervy}},
  \bibinfo {author} {\bibfnamefont {A.}~\bibnamefont {Shalabney}}, \bibinfo
  {author} {\bibfnamefont {E.}~\bibnamefont {Devaux}}, \bibinfo {author}
  {\bibfnamefont {H.}~\bibnamefont {Hiura}}, \bibinfo {author} {\bibfnamefont
  {C.}~\bibnamefont {Genet}},\ and\ \bibinfo {author} {\bibfnamefont {T.~W.}\
  \bibnamefont {Ebbesen}},\ }\bibfield  {title} {\bibinfo {title} {{Multiple
  Rabi Splittings under Ultrastrong Vibrational Coupling}},\ }\href
  {https://doi.org/10.1103/PhysRevLett.117.153601} {\bibfield  {journal}
  {\bibinfo  {journal} {Physical Review Letters}\ }\textbf {\bibinfo {volume}
  {117}},\ \bibinfo {pages} {153601} (\bibinfo {year} {2016})}\BibitemShut
  {NoStop}%
\bibitem [{\citenamefont {Li}\ \emph {et~al.}(2020)\citenamefont {Li},
  \citenamefont {Chen}, \citenamefont {Nitzan},\ and\ \citenamefont
  {Subotnik}}]{Li2020}%
  \BibitemOpen
  \bibfield  {author} {\bibinfo {author} {\bibfnamefont {T.~E.}\ \bibnamefont
  {Li}}, \bibinfo {author} {\bibfnamefont {H.-T.}\ \bibnamefont {Chen}},
  \bibinfo {author} {\bibfnamefont {A.}~\bibnamefont {Nitzan}},\ and\ \bibinfo
  {author} {\bibfnamefont {J.~E.}\ \bibnamefont {Subotnik}},\ }\bibfield
  {title} {\bibinfo {title} {{Quasiclassical modeling of cavity quantum
  electrodynamics}},\ }\href {https://doi.org/10.1103/PhysRevA.101.033831}
  {\bibfield  {journal} {\bibinfo  {journal} {Physical Review A}\ }\textbf
  {\bibinfo {volume} {101}},\ \bibinfo {pages} {033831} (\bibinfo {year}
  {2020})}\BibitemShut {NoStop}%
\bibitem [{\citenamefont {Vasista}\ and\ \citenamefont
  {Barnes}(2020)}]{Vasista2020}%
  \BibitemOpen
  \bibfield  {author} {\bibinfo {author} {\bibfnamefont {A.~B.}\ \bibnamefont
  {Vasista}}\ and\ \bibinfo {author} {\bibfnamefont {W.~L.}\ \bibnamefont
  {Barnes}},\ }\bibfield  {title} {\bibinfo {title} {{Molecular Monolayer
  Strong Coupling in Dielectric Soft Microcavities}},\ }\href
  {https://doi.org/10.1021/acs.nanolett.9b04996} {\bibfield  {journal}
  {\bibinfo  {journal} {Nano Letters}\ }\textbf {\bibinfo {volume} {20}},\
  \bibinfo {pages} {1766} (\bibinfo {year} {2020})}\BibitemShut {NoStop}%
\bibitem [{\citenamefont {Saleh}\ and\ \citenamefont
  {Teich}(1991)}]{Saleh1991}%
  \BibitemOpen
  \bibfield  {author} {\bibinfo {author} {\bibfnamefont {B.~E.~A.}\
  \bibnamefont {Saleh}}\ and\ \bibinfo {author} {\bibfnamefont {M.~C.}\
  \bibnamefont {Teich}},\ }\href {https://doi.org/10.1002/0471213748} {\emph
  {\bibinfo {title} {{Fundamentals of Photonics}}}},\ Wiley Series in Pure and
  Applied Optics\ (\bibinfo  {publisher} {John Wiley {\&} Sons, Inc.},\
  \bibinfo {address} {New York, USA},\ \bibinfo {year} {1991})\BibitemShut
  {NoStop}%
\bibitem [{\citenamefont {T{\"{o}}rm{\"{a}}}\ and\ \citenamefont
  {Barnes}(2015)}]{Torma2015}%
  \BibitemOpen
  \bibfield  {author} {\bibinfo {author} {\bibfnamefont {P.}~\bibnamefont
  {T{\"{o}}rm{\"{a}}}}\ and\ \bibinfo {author} {\bibfnamefont {W.~L.}\
  \bibnamefont {Barnes}},\ }\bibfield  {title} {\bibinfo {title} {{Strong
  coupling between surface plasmon polaritons and emitters: a review}},\ }\href
  {https://doi.org/10.1088/0034-4885/78/1/013901} {\bibfield  {journal}
  {\bibinfo  {journal} {Reports on Progress in Physics}\ }\textbf {\bibinfo
  {volume} {78}},\ \bibinfo {pages} {013901} (\bibinfo {year} {2015})},\
  \Eprint {https://arxiv.org/abs/1405.1661} {1405.1661} \BibitemShut {NoStop}%
\bibitem [{\citenamefont {Agranovich}\ \emph {et~al.}(2003)\citenamefont
  {Agranovich}, \citenamefont {Litinskaia},\ and\ \citenamefont
  {Lidzey}}]{Agranovich2003}%
  \BibitemOpen
  \bibfield  {author} {\bibinfo {author} {\bibfnamefont {V.~M.}\ \bibnamefont
  {Agranovich}}, \bibinfo {author} {\bibfnamefont {M.}~\bibnamefont
  {Litinskaia}},\ and\ \bibinfo {author} {\bibfnamefont {D.~G.}\ \bibnamefont
  {Lidzey}},\ }\bibfield  {title} {\bibinfo {title} {{Cavity polaritons in
  microcavities containing disordered organic semiconductors}},\ }\href
  {https://doi.org/10.1103/PhysRevB.67.085311} {\bibfield  {journal} {\bibinfo
  {journal} {Physical Review B}\ }\textbf {\bibinfo {volume} {67}},\ \bibinfo
  {pages} {085311} (\bibinfo {year} {2003})}\BibitemShut {NoStop}%
\bibitem [{\citenamefont {Bellessa}\ \emph {et~al.}(2004)\citenamefont
  {Bellessa}, \citenamefont {Bonnand}, \citenamefont {Plenet},\ and\
  \citenamefont {Mugnier}}]{Bellessa2004}%
  \BibitemOpen
  \bibfield  {author} {\bibinfo {author} {\bibfnamefont {J.}~\bibnamefont
  {Bellessa}}, \bibinfo {author} {\bibfnamefont {C.}~\bibnamefont {Bonnand}},
  \bibinfo {author} {\bibfnamefont {J.~C.}\ \bibnamefont {Plenet}},\ and\
  \bibinfo {author} {\bibfnamefont {J.}~\bibnamefont {Mugnier}},\ }\bibfield
  {title} {\bibinfo {title} {{Strong Coupling between Surface Plasmons and
  Excitons in an Organic Semiconductor}},\ }\href
  {https://doi.org/10.1103/PhysRevLett.93.036404} {\bibfield  {journal}
  {\bibinfo  {journal} {Physical Review Letters}\ }\textbf {\bibinfo {volume}
  {93}},\ \bibinfo {pages} {36404} (\bibinfo {year} {2004})}\BibitemShut
  {NoStop}%
\bibitem [{\citenamefont {Zengin}\ \emph {et~al.}(2013)\citenamefont {Zengin},
  \citenamefont {Johansson}, \citenamefont {Johansson}, \citenamefont
  {Antosiewicz}, \citenamefont {K{\"{a}}ll},\ and\ \citenamefont
  {Shegai}}]{Zengin2013}%
  \BibitemOpen
  \bibfield  {author} {\bibinfo {author} {\bibfnamefont {G.}~\bibnamefont
  {Zengin}}, \bibinfo {author} {\bibfnamefont {G.}~\bibnamefont {Johansson}},
  \bibinfo {author} {\bibfnamefont {P.}~\bibnamefont {Johansson}}, \bibinfo
  {author} {\bibfnamefont {T.~J.}\ \bibnamefont {Antosiewicz}}, \bibinfo
  {author} {\bibfnamefont {M.}~\bibnamefont {K{\"{a}}ll}},\ and\ \bibinfo
  {author} {\bibfnamefont {T.}~\bibnamefont {Shegai}},\ }\bibfield  {title}
  {\bibinfo {title} {{Approaching the strong coupling limit in single plasmonic
  nanorods interacting with J-aggregates.}},\ }\href
  {https://doi.org/10.1038/srep03074} {\bibfield  {journal} {\bibinfo
  {journal} {Scientific reports}\ }\textbf {\bibinfo {volume} {3}},\ \bibinfo
  {pages} {3074} (\bibinfo {year} {2013})}\BibitemShut {NoStop}%
\bibitem [{\citenamefont {Rozenman}\ \emph {et~al.}(2018)\citenamefont
  {Rozenman}, \citenamefont {Akulov}, \citenamefont {Golombek},\ and\
  \citenamefont {Schwartz}}]{Rozenman2018}%
  \BibitemOpen
  \bibfield  {author} {\bibinfo {author} {\bibfnamefont {G.~G.}\ \bibnamefont
  {Rozenman}}, \bibinfo {author} {\bibfnamefont {K.}~\bibnamefont {Akulov}},
  \bibinfo {author} {\bibfnamefont {A.}~\bibnamefont {Golombek}},\ and\
  \bibinfo {author} {\bibfnamefont {T.}~\bibnamefont {Schwartz}},\ }\bibfield
  {title} {\bibinfo {title} {{Long-Range Transport of Organic
  Exciton-Polaritons Revealed by Ultrafast Microscopy}},\ }\href
  {https://doi.org/10.1021/acsphotonics.7b01332} {\bibfield  {journal}
  {\bibinfo  {journal} {ACS Photonics}\ }\textbf {\bibinfo {volume} {5}},\
  \bibinfo {pages} {105} (\bibinfo {year} {2018})}\BibitemShut {NoStop}%
\bibitem [{\citenamefont {Askenazi}\ \emph {et~al.}(2014)\citenamefont
  {Askenazi}, \citenamefont {Vasanelli}, \citenamefont {Delteil}, \citenamefont
  {Todorov}, \citenamefont {Andreani}, \citenamefont {Beaudoin}, \citenamefont
  {Sagnes},\ and\ \citenamefont {Sirtori}}]{Askenazi2014}%
  \BibitemOpen
  \bibfield  {author} {\bibinfo {author} {\bibfnamefont {B.}~\bibnamefont
  {Askenazi}}, \bibinfo {author} {\bibfnamefont {A.}~\bibnamefont {Vasanelli}},
  \bibinfo {author} {\bibfnamefont {A.}~\bibnamefont {Delteil}}, \bibinfo
  {author} {\bibfnamefont {Y.}~\bibnamefont {Todorov}}, \bibinfo {author}
  {\bibfnamefont {L.~C.}\ \bibnamefont {Andreani}}, \bibinfo {author}
  {\bibfnamefont {G.}~\bibnamefont {Beaudoin}}, \bibinfo {author}
  {\bibfnamefont {I.}~\bibnamefont {Sagnes}},\ and\ \bibinfo {author}
  {\bibfnamefont {C.}~\bibnamefont {Sirtori}},\ }\bibfield  {title} {\bibinfo
  {title} {{Ultra-strong light–matter coupling for designer Reststrahlen
  band}},\ }\href {https://doi.org/10.1088/1367-2630/16/4/043029} {\bibfield
  {journal} {\bibinfo  {journal} {New Journal of Physics}\ }\textbf {\bibinfo
  {volume} {16}},\ \bibinfo {pages} {043029} (\bibinfo {year}
  {2014})}\BibitemShut {NoStop}%
\bibitem [{Note1()}]{Note1}%
  \BibitemOpen
  \bibinfo {note} {For example, in the case of the transition from weak
  coupling to strong coupling, the transition can be described by a
  non-hermitian Hamiltonian which accounts for the loss rates, and these
  completely determine the interaction strength for the transition happens
  (i.e. when the energy degeneracy is removed). The transition we observe here
  is also clearly different from the crossover to the ultra-strong coupling
  regime, as the strong-coupling Hamiltonian is simply an approximate
  representation of the ultrastrong coupling Hamiltonian under the
  rotating-wave approximation. In contrast, since the structures of $H_{N+1}$
  and $H_{2N}$ are completely different, neither one of them can be obtained
  from the other in any limit. Moreover, it can be shown that the behavior seen
  in Fig.~\ref {fig:disp} is reproduced even when a full Hamiltonian, which
  includes the counter-rotating terms (which is typically used under
  ultrastrong coupling conditions), is used (see Supplemental
  Material).}\BibitemShut {Stop}%
\bibitem [{Note2()}]{Note2}%
  \BibitemOpen
  \bibinfo {note} {Since close to the transition neither one of the
  hamiltonians can be used, the length-dependent polariton energies used for
  calculating the polaritonic phase were obtained from the TMM simulations, in
  a similar manner to Fig.~\ref {fig:withL}}\BibitemShut {NoStop}%
\bibitem [{\citenamefont {Zhu}\ \emph {et~al.}(1990)\citenamefont {Zhu},
  \citenamefont {Gauthier}, \citenamefont {Morin}, \citenamefont {Wu},
  \citenamefont {Carmichael},\ and\ \citenamefont {Mossberg}}]{Zhu1990}%
  \BibitemOpen
  \bibfield  {author} {\bibinfo {author} {\bibfnamefont {Y.}~\bibnamefont
  {Zhu}}, \bibinfo {author} {\bibfnamefont {D.~J.}\ \bibnamefont {Gauthier}},
  \bibinfo {author} {\bibfnamefont {S.~E.}\ \bibnamefont {Morin}}, \bibinfo
  {author} {\bibfnamefont {Q.}~\bibnamefont {Wu}}, \bibinfo {author}
  {\bibfnamefont {H.~J.}\ \bibnamefont {Carmichael}},\ and\ \bibinfo {author}
  {\bibfnamefont {T.~W.}\ \bibnamefont {Mossberg}},\ }\bibfield  {title}
  {\bibinfo {title} {{Vacuum Rabi splitting as a feature of linear-dispersion
  theory: Analysis and experimental observations}},\ }\href
  {https://doi.org/10.1103/PhysRevLett.64.2499} {\bibfield  {journal} {\bibinfo
   {journal} {Physical Review Letters}\ }\textbf {\bibinfo {volume} {64}},\
  \bibinfo {pages} {2499} (\bibinfo {year} {1990})}\BibitemShut {NoStop}%
\bibitem [{\citenamefont {Dong}\ \emph {et~al.}(2009)\citenamefont {Dong},
  \citenamefont {Zou}, \citenamefont {Zhang}, \citenamefont {Yang},
  \citenamefont {Li},\ and\ \citenamefont {Guo}}]{Dong2009}%
  \BibitemOpen
  \bibfield  {author} {\bibinfo {author} {\bibfnamefont {Y.-L.}\ \bibnamefont
  {Dong}}, \bibinfo {author} {\bibfnamefont {X.-B.}\ \bibnamefont {Zou}},
  \bibinfo {author} {\bibfnamefont {S.-L.}\ \bibnamefont {Zhang}}, \bibinfo
  {author} {\bibfnamefont {S.}~\bibnamefont {Yang}}, \bibinfo {author}
  {\bibfnamefont {C.-F.}\ \bibnamefont {Li}},\ and\ \bibinfo {author}
  {\bibfnamefont {G.-C.}\ \bibnamefont {Guo}},\ }\bibfield  {title} {\bibinfo
  {title} {{Cavity-QED-based phase gate for photonic qubits}},\ }\href
  {https://doi.org/10.1080/09500340903003339} {\bibfield  {journal} {\bibinfo
  {journal} {Journal of Modern Optics}\ }\textbf {\bibinfo {volume} {56}},\
  \bibinfo {pages} {1230} (\bibinfo {year} {2009})}\BibitemShut {NoStop}%
\bibitem [{\citenamefont {Prado}\ \emph {et~al.}(2011)\citenamefont {Prado},
  \citenamefont {Luiz}, \citenamefont {Villas-B{\^{o}}as}, \citenamefont
  {Alcalde}, \citenamefont {Duzzioni},\ and\ \citenamefont {Sanz}}]{Prado2011}%
  \BibitemOpen
  \bibfield  {author} {\bibinfo {author} {\bibfnamefont {F.~O.}\ \bibnamefont
  {Prado}}, \bibinfo {author} {\bibfnamefont {F.~S.}\ \bibnamefont {Luiz}},
  \bibinfo {author} {\bibfnamefont {J.~M.}\ \bibnamefont {Villas-B{\^{o}}as}},
  \bibinfo {author} {\bibfnamefont {A.~M.}\ \bibnamefont {Alcalde}}, \bibinfo
  {author} {\bibfnamefont {E.~I.}\ \bibnamefont {Duzzioni}},\ and\ \bibinfo
  {author} {\bibfnamefont {L.}~\bibnamefont {Sanz}},\ }\bibfield  {title}
  {\bibinfo {title} {{Atom-mediated effective interactions between modes of a
  bimodal cavity}},\ }\href {https://doi.org/10.1103/PhysRevA.84.053839}
  {\bibfield  {journal} {\bibinfo  {journal} {Physical Review A - Atomic,
  Molecular, and Optical Physics}\ }\textbf {\bibinfo {volume} {84}},\ \bibinfo
  {pages} {053839} (\bibinfo {year} {2011})}\BibitemShut {NoStop}%
\bibitem [{\citenamefont {Wickenbrock}\ \emph {et~al.}(2013)\citenamefont
  {Wickenbrock}, \citenamefont {Hemmerling}, \citenamefont {Robb},
  \citenamefont {Emary},\ and\ \citenamefont {Renzoni}}]{Wickenbrock2013}%
  \BibitemOpen
  \bibfield  {author} {\bibinfo {author} {\bibfnamefont {A.}~\bibnamefont
  {Wickenbrock}}, \bibinfo {author} {\bibfnamefont {M.}~\bibnamefont
  {Hemmerling}}, \bibinfo {author} {\bibfnamefont {G.~R.~M.}\ \bibnamefont
  {Robb}}, \bibinfo {author} {\bibfnamefont {C.}~\bibnamefont {Emary}},\ and\
  \bibinfo {author} {\bibfnamefont {F.}~\bibnamefont {Renzoni}},\ }\bibfield
  {title} {\bibinfo {title} {{Collective strong coupling in multimode cavity
  QED}},\ }\href {https://doi.org/10.1103/PhysRevA.87.043817} {\bibfield
  {journal} {\bibinfo  {journal} {Physical Review A}\ }\textbf {\bibinfo
  {volume} {87}},\ \bibinfo {pages} {043817} (\bibinfo {year}
  {2013})}\BibitemShut {NoStop}%
\end{thebibliography}%

\end{document}